\documentclass[sigconf]{aamas}

\usepackage{balance} 

\usepackage{nicefrac}
\usepackage{cleveref} 
\usepackage{subcaption}

\usepackage{xcolor}
\newcount\Comments 
\definecolor{darkgreen}{rgb}{0,0.6,0} 
\definecolor{union_garnet}{RGB}{134, 38, 51}
 
\newcommand{\kibitz}[2]{\ifnum\Comments=1{\color{#1}{#2}}\fi}

 \setcopyright{rightsretained}
 \acmConference[GAIW'26]{Appears at the 8th Games, Agents, and Incentives Workshop (GAIW-26). Held as part of the Workshops at the 25th International Conference on Autonomous Agents and Multiagent Systems.}{May 2026}{Paphos, Cyprus}{Armstrong, Curry, Hosseini, Mattei, Tsang, Wąs (Chairs)}

\title[Analyzing the Effects of Two-Stage Peer Evaluation]{Analyzing the Effects of Two-Stage Peer Evaluation}

\author{Roy Fairstein}
\affiliation{
 \institution{Ben-Gurion University of the Negev}
 \country{Israel}}
\email{royfa@post.bgu.ac.il}

\author{Harper Lyon}
\affiliation{
 \institution{Tulane University}
 \country{USA}}
\email{hlyon@tulane.edu}

\author{Oshri Damty}
\affiliation{
 \institution{Ben-Gurion University of the Negev}
 \country{Israel}}

\author{Omer Lev}
\affiliation{
 \institution{Ben-Gurion University of the Negev}
 \country{Israel}}
\email{omerlev@bgu.ac.il }

\author{Nicholas Mattei}
\affiliation{
 \institution{Tulane University}
 \country{USA}}
\email{nsmattei@tulane.edu}

\author{Kobi Gal}
\affiliation{
 \institution{Ben-Gurion University and the University of Edinburgh}
 \country{Israel}}
\email{kobig@bgu.ac.il}

\begin{abstract}
    
 \noindent\textbf{Background:} 
 Peer-evaluation and selection systems are used when sets of agents evaluate each other in order to select the best $k$ among them. These are commonly used in real-world settings, including academic conferences where those reviewing papers are often the set of submitters. Conferences have attempted to better allocate their reviewing resources by moving to a two-stage mechanism, in which some papers are eliminated after a first stage of review and remaining papers receive additional reviewers.
    
 \noindent\textbf{Objectives and Research Questions:} 
 We investigate how two major strategyproof peer selection mechanisms, Partition and \emph{ExactDollarPartition}, perform when adapted to a two-stage system, in order to try and understand the effect of the two-stage mechanism on which agents get selected. We also examine how the various parameters of the two-stage mechanism influence the outcome.
    
 \noindent\textbf{Methods:} 
 We provide a theoretical basis by showing how a particular setting is influenced by the two stages. However, solving for the general case seems implausible at the moment, and we use extensive simulations of different scenarios and settings to observe which agents benefit and which are harmed by adopting two-stage mechanisms (and we vary this mechanisms parameters as well).
    
 \noindent\textbf{Results:} 
 We show that the two-stage mechanism's advantage depends the noisiness of reviewer beliefs. Borderline agents benefit most in a low noise environment, while high rank agents benefit more in noisy environments. We show that the effectiveness of these mechanisms is highly dependent on the number of chosen agents, the number of reviews requested from agents, and reviewers' correlation, indicating that organizers need to exercise caution when selecting these parameters for a reviewing process.
    
 \noindent\textbf{Conclusions:}
 We analyze which agents benefit the most from using two-stage mechanisms, and show their value and usefulness beyond the intuition presented so far. Moreover, we are able to improve their performance by choosing appropriate values for the mechanism's parameters.
    
 \end{abstract}


\begin{document}


\pagestyle{fancy}
\fancyhead{}

\frenchspacing
\sloppy
\raggedbottom


\maketitle 


\section{Introduction}
Peer-evaluation, the process in which a group judges its members and selects the best, is a process humans have engaged with for thousands of years. In many ancient cases, selection by lot was done (a form of sortition~\cite{FKP21}), as it was assumed there is a divine intervention to make good selections (see Samuel I, chapter 10). Later, people tried to choose the best person for a job using various selection methods, e.g., the selection of the Pope from -- and by -- the College of Cardinals. While in small groups this can be akin to voting, the main difference from elections is that the set of candidates and voters are the same; and furthermore, the selection of only a single candidate is comparatively rare, more commonly the goal is select a group of the $k$ best agents or to find a ranking of all agents, from which some top-$k$ is typically selected.

More recently, peer selection is commonly practiced in many companies and organizations (sometimes called 360 Evaluations \cite{beehr2001evaluation}), for example, as part of employees' evaluation process or advancement protocol.\footnote{For example, it is a required part of the Israeli Army's officer training course, as well as part of its officer evaluation process.} Academics are intimately familiar with peer selection, as refereed academic conferences, including most computer-science conferences like AAMAS and AAAI, implement a version of such a system: those submitting papers are increasingly often tasked with evaluating other papers \cite{Sha22}. Due to the obvious risks of self interested agents misreporting their evaluations for their own benefit, it is desirable for peer evaluations systems to preempt dishonest agent behavior \cite{gurevych2024reviewer}. As such, research in this field has mostly focused on \emph{strategyproof} mechanisms, i.e., systems in which agents cannot improve their chances of being selected by giving an untruthful review.\footnote{In typical peer-reviewing \emph{other} agents are asked to review, one can view the peer selection problem where these two sets exactly overlap, as a worst case \cite{ALMRW19}. The trend in most major computer science conferences including ICML, NeurIPS, AAMAS, AAAI, IJCAI and others now explicitly \emph{requires} that authors also serve as reviewers, the peer selection setting, with its added challenges, is growing increasingly relevant \cite{goldberg2025peer}.} While we are not implying even the minority of reviewers may be strategic, we take a mechanism design approach and simply want to ensure that the incentives of the system align with the intended outcomes \cite{nisan1999algorithmic}. In addition, these problems are prevalent enough that CVPR\footnote{\url{https://cvpr.thecvf.com/Conferences/2025/CVPRChanges}} and NeurIPS\footnote{\url{https://blog.neurips.cc/2025/05/02/responsible-reviewing-initiative-for-neurips-2025/}} have implemented the specific policy of rejecting author papers if they do not follow the rules when acting as reviewers themselves.

In parallel to these efforts to adjust the review process itself, academic conferences and online courses (MOOCS) \cite{luo2014peer} struggle to deal with the orthogonal problem of how to use a limited resource -- the number of papers that can be reviewed by agents --  to select the best set of papers. As research has shown~\cite{Law22}, reviewers are  ``noisy'', i.e., not in agreement with each other on a ground-truth of the ranking of all papers.\footnote{While we fully agree that not all academic work follows the typical Condorcet Noise Model, where an unobservable ground truth ranking exists, but is only observable through noisy votes \cite{conitzer2005common}, it is a useful model for study. This model provides a baseline to measure against and while stylized, has been extensively used in the peer-reviewing and peer-selection literature. We are not implying that reviewers are intentionally noisy or adversarial, we simply use the setting to gain insights to what happens when we assume that not every reviewer is as reliable as all the others \cite{noothigattu2021loss,wang2019your}.} Therefore there is a desire to get more eyes on each paper so as to increase the chance of getting a better signal regarding its quality. One approach to deal with this issue, initially used at the AAAI 2021 conference, and since then more widely in AI conferences \cite{shah2018design}, is two-stage reviewing~\cite{LMNZCNR22}. In the first stage, each reviewer reviews only a few papers, and based on this signal, papers which received very bad reviews are eliminated. This leaves fewer papers, so that each can receive more reviews than previously possible. Thus, instead of all papers receiving the same amount of reviews, those eliminated in the first stage receive fewer, while the remaining get more, ideally allowing for better selection of the best papers.  

Note that two-staged mechanisms are not just beneficial for paper reviews. They help peer evaluation whenever the evaluating agents are noisy, but at least somewhat positively correlated with the optimal outcome. This assumption holds for many uses of peer evaluation, from MOOC course grading \cite{luo2014peer} to verifying online workers \cite{burmania2015increasing} to employee evaluation \cite{beehr2001evaluation}.

\paragraph{Contribution.} We formally analyses two-stage variants of three peer   selection mechanisms, a standard non-strategyproof \emph{Borda} based mechanism (``vanilla''); \emph{Partition}, an oft-researched strategyproof mechanism~\cite{AFPT11}; and \emph{Exact Dollar Partition}, a mechanism using some Partition ideas, but which has been shown to out-perform it empirically~\cite{ALMRW19}. We examine the following related questions:
\begin{enumerate}
\item What is the quality of a two-stage peer evaluation mechanism? Does it improve over its single stage variant, and where do these improvements manifest themselves? Specifically, which papers benefit from two-stage models, and which may be hurt?
\item What parameters of two-stage mechanisms lead to better results? For instance, if each reviewer reviews, overall, $m$ agents, how many of those should be done in the first round and how many in the second round? 
\item Are the ideal parameters shared between distinct mechanisms, or are they mechanism dependent?
\end{enumerate}
Using extensive simulations, we examine these questions in settings that vary the number of agents we wish to choose ($k$), the number of reviews from each agent ($m$), and the level of noise in agent's preferences. We provide a theoretical justification for our empirical findings that, in general, papers very close to the ``cutoff'' of selection are the most helped (or hurt) by these mechanisms. We also see that the different parameters of the two-stage mechanism are highly influenced by the number to be selected $k$, the number of reviews provided per agent $m$, and the correlation of views, and we detail their connection. 


\section{Related Work}\label{relatedWork}

The AI community's attention to peer-evaluation was sparked by \citet{MS09} who suggested to use  peer-evaluation to divide telescope time. That mechanism incentivized agents not to report their true views of others, but instead, to report as their view what they think the consensus opinion will be. This undesirable property (encouraging misreporting) prodded researchers to suggest strategyproof mechanisms, in which agents are never better off by misreporting their true beliefs on others~\cite{Sha22}.

One prominent mechanism, suggested in \citet{AFPT11}, is Partition, in which agents are divided into two groups which review each other, thus preserving strategyproofness; as each reviewer can only influence the ranking of agents in a group it is not a member. In the original paper, as well as some further ones~\cite{HM13a}, the setting considered was that of nomination, i.e., agents either approve or disapprove of a paper, so no ranking or more granular grades are elicited. Furthermore, some papers focused only on the selection of a single agent ($k=1$)~\cite{HM13a,FK15,BNV14}, though a few expanded beyond that~\cite{BFK17}, including with real-world data~\cite{XZSS19}. A variant of Partition with more than two clusters and different weights to each group, \emph{Dollar Partition}, was suggested in \citet{ALMRW16}, though it could return fewer than $k$ agents, an issue fixed in \emph{Exact Dollar Partition}~\cite{ALMRW19}.

Beyond Partition, several other mechanisms have been suggested. \citet{KLMP15} suggested \emph{Credible Subset}, a mechanism based on identifying agents with a potential to manipulate, and including them in the set of potential winners, though that mechanism had a probability of selecting no agents at all. \citet{GWL19} suggested using a low-probability event of verifying reviewers, punishable very heavily to encourage agents to report truthfully. \citet{MTZ20} proposed \emph{PeerNomination}, which reverts to a nomination (or approval) mechanism, in which agents either approve or disapprove of a nomination, though that mechanism suffers from a possibility of returning fewer agents than required. In order to tackle potentially malicious (or just bad) reviewers, \citet{LMTZ23} reworked PeerNomination so it weighs down suspected low-quality reviewers, while maintaining strategyproofness. On the other hand, \citet{Wal14} suggested a PageRank-like peer-evaluation method, that gives up on strategyproofness in order to be able to weigh down less-regarded reviewers. In addition, several algorithms have been suggested involving how to deal with various biases and issues in agents' bidding and grading-normalization~\cite{NSP19,SSS19,WS19}. Additional discussions can be found in the annotated reading list of \citet{lev2024impartial} and the survey by \citet{olckers2024manipulation}.

To model the (potential) inaccuracies in reviewers’ assessments, we assume that each agent is associated with a noisy observation of the ground truth according to a Mallows model~\cite{Mal57}. These have been widely used to compare peer-evaluation mechanisms~\cite{ALMRW16,ALMRW19,MTZ20,LMTZ23}. This is often called the Condorcet Noise Model in the literature \cite{conitzer2005common}. While this model does not fully capture reality, it has also been extensively used to highlight and study problems of calibration and other issues in peer review and evaluation \cite{noothigattu2021loss,wang2019your}.

As discussed in the first section, due to the incredible increase in submissions, and hence, reviews, required at large AI conferences, studying mechanisms to improve the peer review process has become an important topic of study broadly in AI \cite{shah2018design}, see \citet{Sha22} for a comprehensive survey. We focus on the question of how to allocate our (often scarce) reviewing resources, which also borrows from the extensive literature on artifact verification from the crowd-sourcing literature \cite{burmania2015increasing,baba2013statistical}. Recent papers have called for more incentives, and a closer investigation of what practices work (and which do not) in the peer review space \cite{kim2025positionaiconferencepeer,gurevych2024reviewer,goldberg2025peer}.

\section{Preliminaries}\label{prelim}

In our setting, peer-evaluation agents will be denoted by $N=\{1,2,\ldots,n\}$. The agents in peer-evaluation represent both candidates that wish to be selected, as well as agents which vote on the selection by ranking candidates. Of the $n$ candidates, we wish to select a subset of size $k$.

In this paper we will assume agents give grades to the candidates that they review\footnote{As noted above, other possibilities in the literature include approval/disapproval or rankings. Of course, numerical grades can be easily translated into a ranking.}. Each agent $i$ has a set $N_{i}\subseteq N$ of $m$ agents that they review and grade, i.e., each agent has a function $A_{i}:N_{i}\rightarrow \mathbb{R}$, the grade it gives each paper.

As in previous peer-evaluation papers, we will assume the Condorcet Noise Model and that there is a ground-truth, which is the ranking of the papers and we aim to select the best from that set. For simplicity, we assume the agents are numbered according to their true ranking, so the top $k$ agents are agents $\{1, 2,\ldots,k\}$.

We use the Mallows model \cite{Mal57} to reflect the noisy observation of the ground truth of each agent.\footnote{While there are many models and guides for election related experiments \cite{boehmer2024guide} we focus on the Mallows model for its simplicity and to guide later investigation into more realistic and potentially real-world data and experiments \cite{mattei2013preflib}.} The Mallows model is parameterized by a dispersion $\phi\in [0,1]$ and the ground truth ranking $\sigma\in\pi(N)$, for $\pi(N)$ being all possible orderings of $N$ agents. 
The Kendall-$\tau$ distance counts the number of pairwise disagreements between two rankings, and for any two rankings of the same $N$ agents $a,b\in\pi(N)$, the value $d(a,b)$ shall denote the Kendall-$\tau$ distance between them \cite{kendall1938new}. This is also sometimes called the bubble-sort distance as it is the number of distinct swaps one must make to convert one ranking into another. The Mallows model first builds a probability space from which agents' ranking are sampled. For any ranking $r\in\pi(N)$, the probability of an agent having that ranking is $P(r) = P(r \mid \sigma,\phi)=\frac{1}{Z}$ $\phi^{d(r,\sigma)}$, where $Z = 1\cdot(1+\phi)\cdot(1+\phi+\phi^2) \ldots (1+ \phi+ \ldots +\phi^{n-1})$. Informally, this is an exponential distribution where the probability of pairwise swaps increases as agents become more noisy. For every agent $i\in N$, we select their own noisy ranking from this probability space.\footnote{To convert from each agent's ranking to its grade, we gave the top candidate 100, and the grades decrease, point by point, as a candidate is located lower in the ranking. The function $A_{i}$ returns the grade, based on this ranking. This Borda-like score has been used in other peer-evaluation papers~\cite{ALMRW16,ALMRW19}.}

\subsection{Partition Algorithm}
The Partition algorithm, as presented in \citet{AFPT11}, takes the set of agents, divides them into $c$ clusters, and each agent reviews agents from a set that it is not a member of. From each partition the top $\nicefrac{k}{c}$ agents are selected.

While the original version of the algorithm~\cite{AFPT11} assumed all agents of each partition review \emph{all} agents of the other partition, this is not practical in cases with a large number of agents. Therefore, we assume the existence of a function that assigns agents from one cluster to review $m$ of the other cluster and vice versa (any algorithm that does this can be used, as is assumed in \citet{ALMRW16,ALMRW19}; this can be done greedily, or, as detailed in \citet{LMTZ23}, based on Euler cycles). Note that while we assume (as is common in papers analyzing Partition) that clusters are created randomly, there are also cases where clusters are constructed to satisfy other criteria such as avoiding conflict of interest.

\subsection{Exact Dollar Partition}
Exact Dollar Partition (EDP) attempts to reduce the effects of the allocation randomization in Partition, as it allows, for example, all the best papers to be allocated to the same partition, resulting in only a small part of them being selected by the algorithm. Similarly, if many bad papers are in the same partition, Partition will still select a meaningful part of them as part of its solution.

Exact Dollar Partition solves this by having each agent allocate a fixed number of points (e.g., each agent divides 100 points between all the papers they review), thus, the number of points in the mechanism is fixed (in our example, $100n$). Now, Exact Dollar Partition uses the number of points allocated to each partition as a proxy to the quality of the agents it contains, and choses from each partition only number of agents that it ``deserves'' to get based on its share of the overall score. So if a partition has many good papers, many papers will be selected from it, and if has only bad papers, very few will be selected from it. The original version of the algorithm~\cite{ALMRW16} struggled with fractional partition shares, but later work~\cite{ALMRW19} added a novel allocation algorithm that randomly rounds shares while preserving, in expectation, each partition's share.

\section{Theoretical Basis}\label{sec:theory}

To build a basic understanding on which item would gain most from having additional samples (i.e., in our running example, reviews), we look at a simplified model. We will see that it will show our basic empirical finding -- two-stages help borderline papers in less noisy settings, and help more highly ranked papers in more noisy settings.

We assume there is a probability of $p_{i}$ that a reviewer supports acceptance of paper $i$, and for notation ease assume $p_{1}\geq p_{2}\geq\ldots\geq p_{n}$. We further assume reviewers are independent of each other (and so are their reviews), so the number of accepting votes for paper $i$ following $m$ reviews is distributed binomially: $V_{i}\sim \mathrm{Binomial}(m,p_i)$. The ``score'' for each paper is their average: $\hat{p_{i}}=\frac{V_{i}}{m}$.

We shall now make an assumption that holds only for large sets of reviewers, i.e., values of $m$, but simplifies the following calculations significantly: we use the Central Limit theorem to approximate $\hat{p_{i}}\approx \mathcal N(p_i,\frac{p_i(1-p_i)}{m})$. Furthermore, we assume there is some value $t$, that does not change with $m$ such that if $\hat{p_{i}}\geq t$ it is in the top-$k$ (and thus accepted), and if $\hat{p_{i}}<t$ that $i$ will not be accepted. For notational simplicity we will define $b_{i}=\sqrt{p_{i}(1-p_{i})}$.

This allows us to define a function that depends on the review size -- $m$:
\small
$$P_{i}(m)=P(\hat{p_{i}}\geq t)\approx  1-\Phi(\frac{t-p_i}{\frac{b_i}{\sqrt{m}}})$$
\normalsize
where $\Phi$ denotes the Normal distribution's CDF (cumulative distribution function).

We are interested in $\Delta P_{i}=P_{i}(m+x)-P_{i}(m)$. This is approximated, from the above definitions to $-\Phi(\frac{t-p_i}{\frac{b_i}{\sqrt{m+x}}})+\Phi(\frac{t-p_i}{\frac{b_i}{\sqrt{m}}})=\Phi(\frac{p_i-t}{\frac{b_i}{\sqrt{m+x}}})-\Phi(\frac{p_i-t}{\frac{b_i}{\sqrt{m}}})$. For notational ease, we denote $z_{i}(m)=\frac{(p_{i}-t)\sqrt{m}}{b_{i}}$, we have:
\small
$$\Delta P_{i}\approx \Phi(z_{i}(m+x))-\Phi(z_{i}(m))$$
\normalsize

Thanks to the mean-value theorem, we know there is a point $a\in [z_{i}(m)$ and $z_{i}(m+x)]$ such that $\Delta P_{i}=\varphi(a)(z_{i}(m+x)-z_{i}(m))$, for $\varphi$ the Normal distribution's PDF (Probability Density Function). Putting in the values for $z_{i}$, we end up with
\small
$$\Delta P_{i}\approx \varphi(a)\frac{p_{i}-t}{b_{i}}(\sqrt{m+x}-\sqrt{m})$$
\normalsize
Showing us (since $\varphi$ is never negative) the rather obvious point that if $p_{i}>t$ it stands to benefit from additional reviewers, and if $p_{i}<t$ you will lose from this.

A further simplifying assumption, to provide us with better understanding, is to assume $a$ is the mid-point of the $[z_{i}(m),z_{i}(m+x)]$ range. Due to space constraints, we will not show here the full arithmetic, but the calculation ends up as:
\small
$$\Delta P_{i}\approx (\sqrt{m+x}-\sqrt{m})\frac{p_{i}-t}{b_{i}}\varphi(\frac{p_{i}-t}{b_{i}}\frac{\sqrt{m+x}+\sqrt{m}}{2})$$
\normalsize

Trying to find a value for $\frac{p_{i}-t}{b_{i}}$ that maximizes $\Delta P_{i}$, the $(\sqrt{m+x}-\sqrt{m})$ part can be ignored, but  looking what zeroes the first derivative, we find it is $\frac{p_{i}-t}{b_{i}}=\frac{2}{\sqrt{m+x}+\sqrt{m}}$ (and the second derivative is concave). Replacing $b_{i}$ with its definition we end up with the $i$ that maximizes $\Delta P_{i}$ is such that
\small
$$\frac{p_{i}-t}{\sqrt{p_{i}(1-p_{i})}}=\frac{2}{\sqrt{m+x}+\sqrt{m}}$$
\normalsize

\begin{description}
\item[Low Noise] In such settings, $p_{i}$ is close to 1 for most $i<k$ (thus $\frac{p_{i}-t}{\sqrt{p_{i}(1-p_{i})}}$ is very large) and close to 0 for most $i>k$, and has a a different value only around $k$. This means the only values for which $\frac{p_{i}-t}{\sqrt{p_{i}(1-p_{i})}}$ will approach the desired maximum are $k$ or slightly below it.
\item[High Noise] In such settings, $p_{i}$ is close to being equal amongst all $i$s, as well as $p_{i}(1-p_{i})$ close to being equal. This means the value inside $\varphi$ is close to equal for all $i$, making maximizing $\Delta P_{i}$ basically increasing $p_{i}-t$, thus it is maximized for $p_{1}$.
\item[Intermediate Noise] $\frac{p_{i}-t}{\sqrt{p_{i}(1-p_{i})}}$ is continuous, and the $i$ maximizing $\Delta P_{i}$ is basically the first index close to $\frac{2}{\sqrt{m+x}+\sqrt{m}}$. As this is continuous, it can be seen to increase from the low-noise case monotonically to the large-noise case.
\end{description}

\section{Methodology}
We ran 3 different mechanisms. In all of them we had $N$ agents, each of them reviewing $m$ agents, with the goal to select $k$ agents. The mechanisms we examined:
\begin{description}
\item[Vanilla] Straightforward, non-strategyproof, mechanism, in which agents rank others using a Borda score, and the $k$ top scoring agents are selected.
\item [Partition] As explained above, agents are randomly assigned to two clusters, with each agent reviewing only agents from the other cluster. The top $\nicefrac{k}{2}$ agents of each cluster are selected~\cite{AFPT11}.
\item[Exact Dollar Partition] As noted above, agents are randomly assigned to one of 3 clusters, with each agent reviewing only agents that are not in their own cluster. Agents' score is normalized (so each contributes the same number of points in the mechanism), and the share of the score given to each cluster determines how many agents are selected from the cluster (i.e. if one cluster of the three received half of the score given by reviewers, $\nicefrac{k}{2}$ agents will be selected from it, even though the ``equal share'' is $\nicefrac{k}{3}$)~\cite{ALMRW19}.
\end{description}

We assess the accuracy of the mechanism using three key metrics. The first is the widely used precision@$k$ metric~\citep{sujatha2011precision}, commonly employed in information retrieval and recommendation systems \cite{ricci2021recommender}. This metric indicates the proportion of agents correctly identified within the top-$k$ positions of the ground truth ranking.

While precision@$k$ is a standard measure, it has a notable limitation: it does not differentiate between which agents were selected. As a result, the score remains the same whether the mechanism fails to identify the agent ranked first or the agent ranked at the $k$-th position.

To address this limitation, we examined two additional metrics:
\begin{description}
 \item [Positive Borda] The top-ranked agent is assigned a score of $k$, with the score gradually decreasing until the agent ranked $k$ receives a score of one; all others receive zero. The Positive Borda score is calculated as the ratio of the score of the selected agents to the optimal score.

 \item [Negative Borda] Similar to Positive Borda, but here the $n$-ranked agent in the ground truth receives the highest score, and the agent ranked $k+1$ receives a score of one.
\end{description}

The Positive Borda score helps us assess how highly ranked the omitted agents are according to the ground truth, while the Negative Borda score indicates how low the incorrectly selected agents rank. In practice, making extreme errors is unlikely, leading to a high correlation between these metrics and precision@$k$. Consequently, for the remainder of the paper, we will focus primarily on the precision@$k$ metric, mentioning Positive and Negative Borda when they are instructive in gaining insight into the algorithm function; but readers should be aware that similar results were obtained for these metrics as well. 

We examine various values of the possible parameters involved in peer evaluation generally, as well as particular values relevant to two-stage mechanisms.  
\begin{description}
\item [$\phi$] Mallows dispersion parameter. Tested values: 0.2, 0.5, 0.8, 0.95.
\item [$k$] Number of chosen agents. Tested values: 10, 20, 30, 40.
\item [$m$] Number of reviews per agent. Tested values: 5, 7, 15.
\item [$f$] Number of reviews in first round. Tested values: $\frac{1m}{10}$,$\frac{2m}{10}$,$\frac{3m}{10}$,$\frac{4m}{10}$.
\item [$h$] Size of the higher candidates group (i.e., is chosen after the first round). Tested values: $0$,$\frac{k}{10}$,$\frac{2k}{10}$,$\frac{3k}{10}$,$\frac{4k}{10}$,$\frac{k}{2}$.
\item [$l$] Size of the lower candidates group (i.e., is eliminated after the first round). Tested values: $100, 125, 150$.
\end{description}

The experiment was repeated $10,000$ times for each setting. We compared all algorithms in each setting with the same clusters to be able to compare directly the effects of the algorithm and not the random assignment into clusters. 

\section{Results}

In this section we will first look at each of the selection algorithms in turn, as there are important differences in their selection behavior. After this, we turn to understanding the overall outcome quality across the mechanisms.

\subsection{Selection Mechanism Results}

\subsubsection{Vanilla Model}
\begin{figure}[ht!]
\begin{center}
\includegraphics[width=0.7\linewidth]{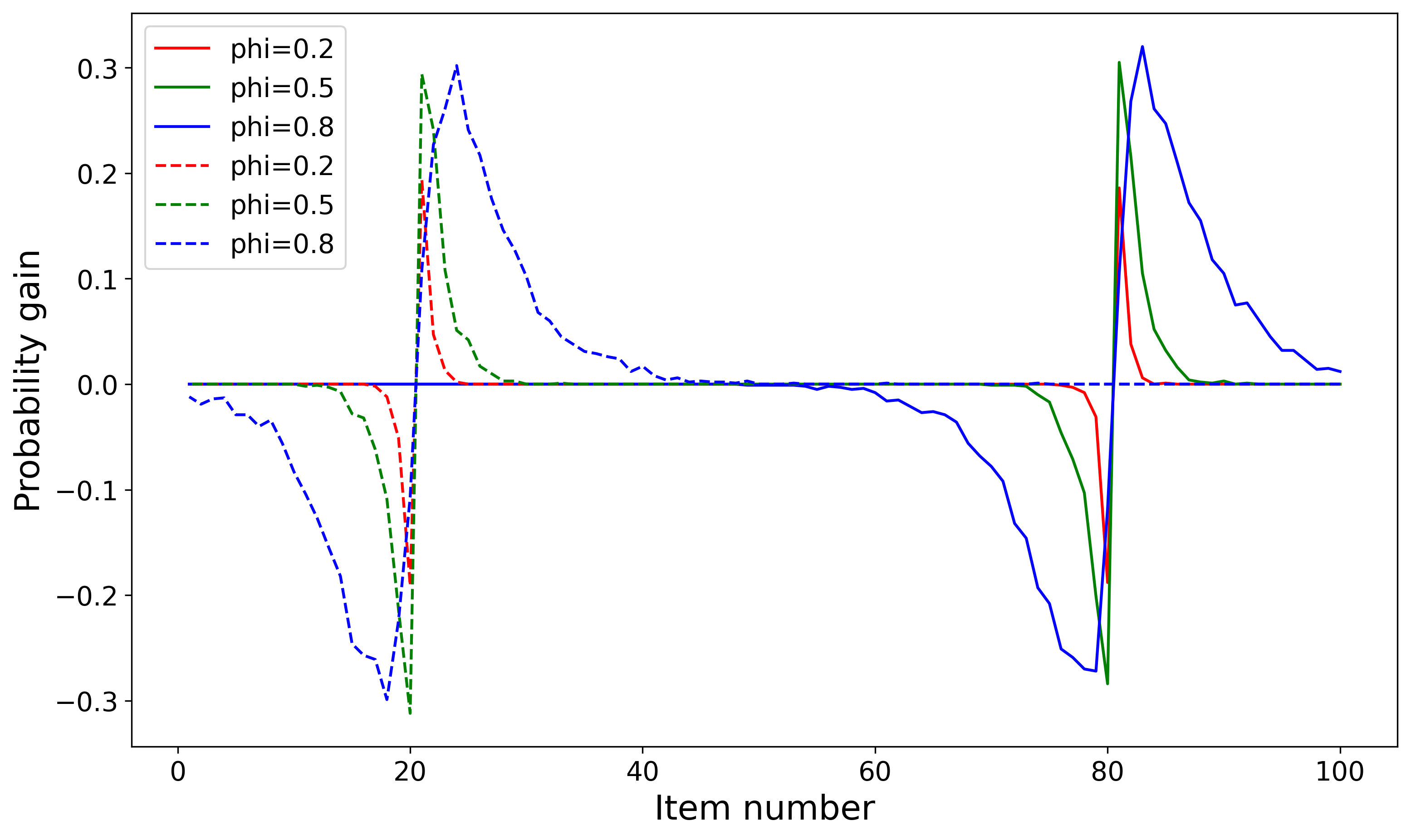}
\caption{Items gain in probability for different amounts of noise when selecting 20 items (full line) and selecting 80 items (dashed line).
}\label{fig:vanilla}
\Description{Graph presenting the gain per agent in probability of selection at differing noise levels.}
\end{center}
\end{figure}

Figure~\ref{fig:vanilla}, detailing a setting with $N=100$, shows a gain by each item of being selected in the top $k$ by having 20 reviews vs. the baseline of a single review. The solid lines show the gains when selecting the top 20 items; and the dashed lines show the gains when selecting the top 80 items.

The results mirror the intuition presented in the previous section: With a small amount of noise (i.e., a low $\phi$ value), the $k$-th and $k+1$ items gain / lose most due to the added samples, and the items further away are barely affected. But as the noise increases (i.e., $\phi$ grows), we see an increase in the gain size itself -- the contribution of the additional information is larger -- and the affected agents change: the location of the peak is further away from the $k$-th item. That is, the item increasing its chance of being selected in the top $k$ is less than $k$, and the item whose probability of being selected drops most significantly is more than $k+1$. This stems, as noted before, from the borderline around $k$ being unclear, due to the noise level, even with more samples; in extremis, when $\phi=1$, the added samples would not help at all.

Two more phenomena can be seen in Figure~\ref{fig:vanilla}. First, we see the gains / loses around the $k$-th items being a mirror image of each other, such that they have similar loses / gains at the same distance from $k$ (with different signs, of course). Second, we see the gains when selecting the top 20 or top 80, which look almost the same. Therefore, we surmise that the items that gain the most -- and how many are affected at all -- mainly depend on the noise and the distance from the borderline value ($k$), and not on how many items are selected.

\subsubsection{Strategyproof Mechanisms}
Similar to vanilla, we wish to see how the strategyproof mechanisms, specifically Partition and Exact Dollar Partition, change with two-stage mechanisms. Both mechanisms are based on partitioning the reviewers into clusters, so that each reviewer can review only others from a different cluster. In the case of partition, the exact number of items are selected from each cluster. In contrast, Exact Dollar Partition weights the clusters according to the reviews, and then the number of items from each cluster is chosen according to these weights.

Crucially, in both of these mechanisms the outcome depends heavily on the specific clusters chosen. A particular division of clusters may give us the ground truth, while another will cause some items to never be selected, even if they should be. A simple such example is Partition with two clusters. Unless exactly half of the top-$k$ items appear in each cluster, it will never be possible to choose all of the top-$k$ items.

Therefore, looking at the general gains of each item like we did in Figure~\ref{fig:vanilla} can be more confusing than helpful -- it is better to look at each cluster separately. Looking at each cluster, we basically get the Vanilla mechanism, i.e., a set of agents, from which a fixed number is selected. We might expect similar gains as shown in Figure~\ref{fig:vanilla}, but in contrast to Vanilla, each cluster is no longer continuous, that is, composed of the full rank of agents -- both the agent ranked $i$ and that ranked $i+1$. This means that as we have more clusters, the gap between two consecutive items in some clusters is more likely to be more significant. For example, an item ranked $i$ is followed by an item ranked $i+3$ -- and the probability of an agent ranking $i+3$ above $i$ is smaller than that of an agent ranking $i+1$ above $i$. Thus, unlike Vanilla, it requires more noise to actually confuse two continuous items in a cluster, which means having more clusters mitigates the effects caused by noise, making the benefit of requesting more reviews redundant.

\begin{figure}[!ht]
\begin{center}
\includegraphics[width=0.7\linewidth]{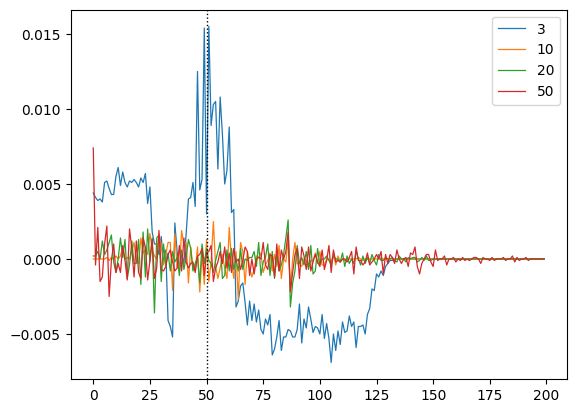}
\caption{Items gain in probability for different amounts of clusters when selecting 50 items from 20 samples compared to 1 sample, with $\phi=0.2$ and using partition for aggregation.
}
\label{fig:clusters_partition}
\Description{Graph presenting effect of cluster size on a two stage peer selection instance.}
\end{center}
\end{figure}

\subsubsection{Varying The Number of Clusters}
This raises a trade-off: when using more clusters, fewer reviews are needed to determine the ordering inside each one, but the outcome is much more dependent on the allocation of the agents between the clusters. Of course, Exact Dollar Partition was designed specifically to mitigate this (and as will be seen below, it often succeeds), but as it is also stochastic, we might "throw away" some of the top-$k$ items.

To examine this, we ran Partition and Exact Dollar Partition with a variety of cluster numbers: $3,10,20,50$\footnote{Other variable values were $N=200$ and $k=50$.}. For each of 10,000 experiments we sampled different profiles and clusters, and calculated the average gain of moving from 1 sample to 20 samples (the exact triplets of profile, clusters, and assignments were tested for all cluster sizes).

Figure~\ref{fig:clusters_partition} shows the effects of adding reviews with Partition. As expected, when having only 3 clusters, we see that items can gain from taking extra samples. However, as we increase the number of clusters, we see that the advantage from additional reviews becomes very close to zero and looks like noise due to the choice of the clusters.

\begin{figure}[ht!]
\begin{center}
\includegraphics[width=0.7\linewidth]{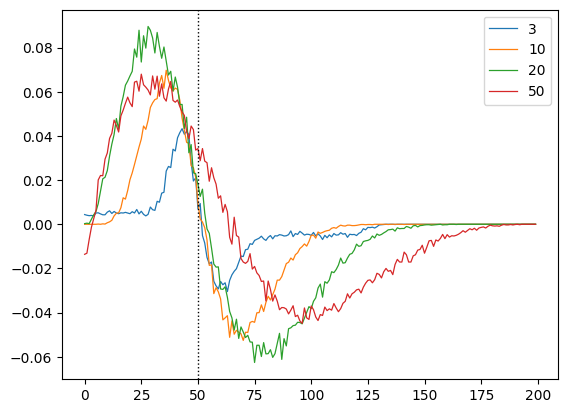}
\caption{Items gain in probability for different amounts of clusters when selecting 50 items from 20 samples compared to 1 sample, with $\phi=0.2$ and using partition for Exact Dollar Partition.
}
\label{fig:clusters_edp}
\Description{Graph presenting effect of cluster size on a two stage peer selection instance.}
\end{center}
\end{figure}

Repeating the experiments with Exact Dollar Partition shows very different results, as seen in Figure~\ref{fig:clusters_edp}. As we increase the number of clusters, the improvement in adding the samples increases as well, and the peak moves further away from the $k$-th item (as with vanilla). However, when moving from 20 clusters to 50 clusters, we see the gain is decreasing; but the trend moving away from the $k$-th item, and towards a longer "tail" -- affecting more agents -- continues. 

The growing effect on more agents as the number of clusters increases (in both Partition and Exact Dollar Partition) has to do, we believe, with clusters creating the potential for very low-ranked agents to be selected, and thus even agents far away from the top-50 are affected by the added reviews. Furthermore, these results indicate that while any number of clusters benefits from adding more reviews, there is a "sweet spot" where items have the most to gain. As we show next, this can change depending on the amount of noise. 

The difference in results between Partition and Exact Dollar Partition, despite both basically having clusters, each behaving similar to Vanilla, is due to the agent grade normalization in Exact Dollar Partition. It means that reviewers who are reviewing only good agents (or only bad agents) give them all medium grades. The more reviews an agent does, the less likely such cases become, and normalization becomes less of an issue; but for small reviewing batch (i.e., $m$), this creates noise. In extremis, if each agent reviewed only a single paper, normalization would make all agents' grades equal.

\begin{figure}[ht!]
\begin{center}
\includegraphics[width=0.7\linewidth]{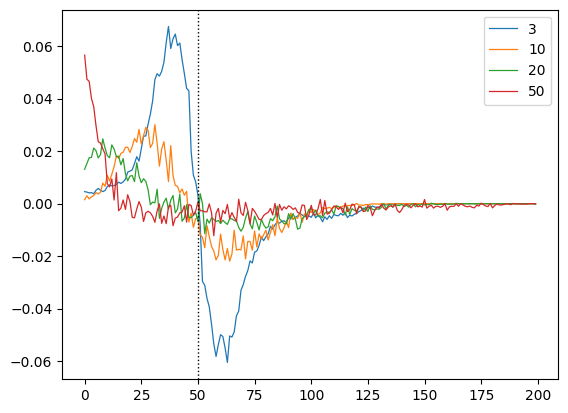}
\caption{Items gain in probability for different amounts of clusters when selecting 50 items from 20 samples compared to 1 sample, with $\phi=0.8$ and using partition for aggregation.
}
\label{fig:clusters_partition08}
\Description{Graph presenting effect of cluster size on a two stage peer selection instance.}
\end{center}
\end{figure}

\begin{figure}[ht!]
\begin{center}
\includegraphics[width=0.7\linewidth]{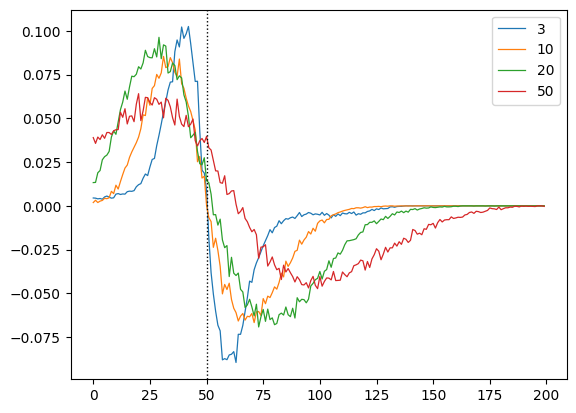}
\caption{Items gain in probability for different amounts of clusters when selecting 50 items from 20 samples compared to 1 sample, with $\phi=0.8$ and using partition for Exact Dollar Partition.
}\label{fig:clusters_edp08}
\Description{Graph presenting effect of cluster size on a two stage peer selection instance.}
\end{center}
\end{figure}

In \Cref{fig:clusters_partition08,fig:clusters_edp08}, we repeated this in a noisier setting. As noise increases, adding reviews is more helpful and significant. Now, even for Partition, with many clusters there is still meaningful improvement by adding reviews. Interestingly, in Exact Dollar Partition there is no monotonically decreasing/increasing improvement as the number of clusters grows (as shown for less noisy settings).

\begin{figure}[ht]
 \centering
 \begin{subfigure}{0.4\textwidth}
 \includegraphics[width=0.7\linewidth]{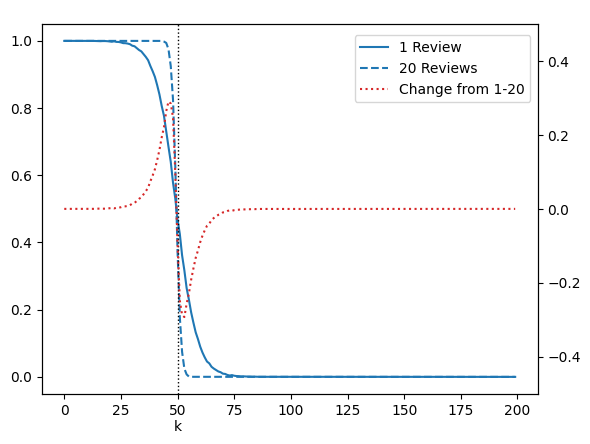}
 \end{subfigure}
 \begin{subfigure}{0.4\textwidth}
 \includegraphics[width=0.7\linewidth]{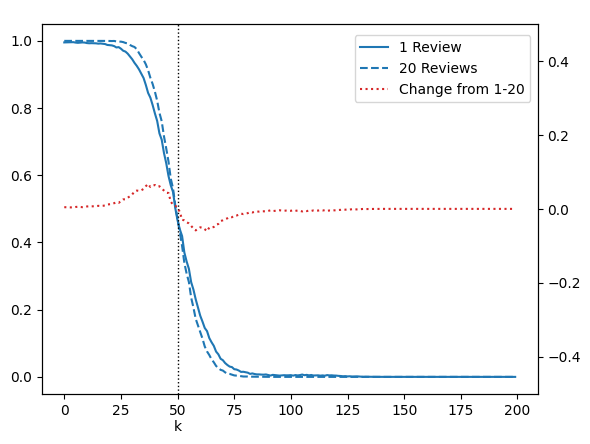}
 \end{subfigure}
 \begin{subfigure}{0.4\textwidth}
 \includegraphics[width=0.7\linewidth]{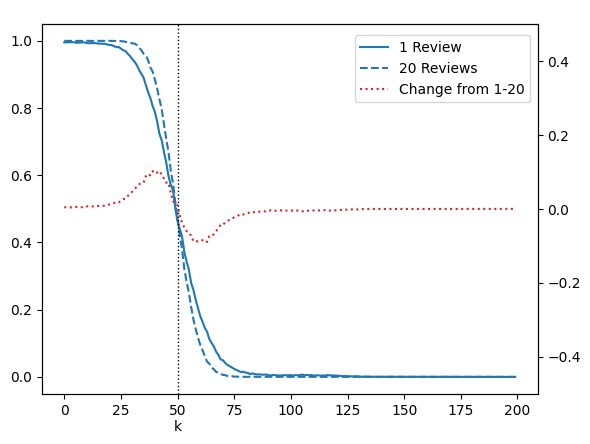}
 \end{subfigure}
 \caption{Vanilla (top), 3 cluster Partition (middle) and 3 cluster Exact Dollar Partition (bottom) agents' probabilities of being selected to the top $k$, with values on the left-hand side. The right-hand side is the values for the delta between a single review and 20 reviews (the red dotted line).}\label{fig:triple_prob_gain}
\end{figure}

Finally, to get a better picture of the competence of these mechanisms, and the effect of adding reviews, \Cref{fig:triple_prob_gain} shows (solid vs. dotted blue line), how adding reviews improves the likelihood of accepting some agents (and rejecting others) in the standard 3 cluster setting. Note how Vanilla (top) improves most dramatically when adding reviewers, while both partition based mechanisms (bottom two) shifts in a much more limited amount. Exact Dollar Partition (bottom) improves on Partition (middle), though does not reach vanilla's magnitude.

\begin{figure}[ht]
\begin{center}
\includegraphics[width=0.6\linewidth]{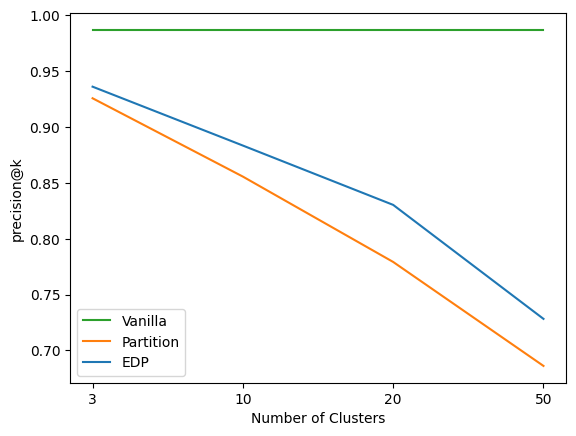}
\includegraphics[width=0.6\linewidth]{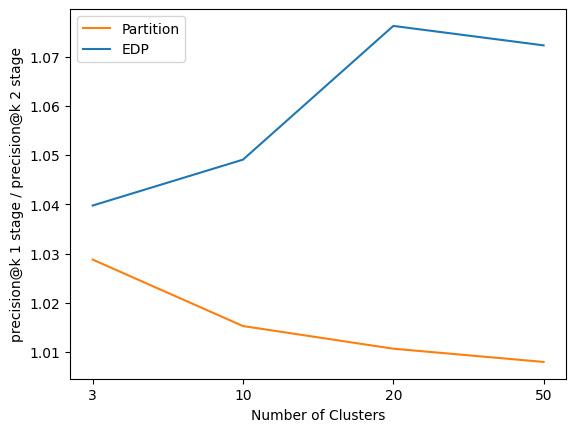}
\caption{Top: precision@k for 200 items taking top 50 with 20 samples and $\phi=0.8$. Bottom: precision@k gains for same parameters.
}\label{fig:prec}
\Description{Graph presenting precision@k and change in precision@k for for a two stage peer selection instance.}
\end{center}
\end{figure}
\subsection{Outcome Quality}


Having established that adding reviews adds to the quality of the outcome, particularly for agents around $k$, we wish to examine various parameters of the mechanism and see which maximize the performance of the two-stage mechanism, both for Vanilla as well as for our two strategyproof mechanisms. Due to real-life reviewing not having, often, a high degree of correlation, we focus on simulations with $\phi=0.95$ -- a rather high value, resulting in meaningful disagreement between reviewers, though still grounded, on a large scale, in the ground truth \cite{boehmer2023properties}.

As shown in our simulations, as well as previous ones~\cite{ALMRW16,ALMRW19}, Vanilla with the (improbable) assumption that all agents are truthful performs better than specifically strategyproof mechanisms, which trade some of the value and flexibility away in order to become strategyproof. However, as will be seen below, using two-stage mechanisms improves performance (and, in some case, allows them to outperform single stage Vanilla, see \Cref{fig:init_rev}). \Cref{fig:prec} shows the Vanilla model performs very well, and it is not surprising to see that as the number of items >> $k$, Partition and Exact Dollar Partition are less likely to choose extremely bad items. Though we see the deterioration of the outcome quality as the number of clusters increase (as noted before).

When looking at the number of clusters we can see opposite results for Exact Dollar Partition and Partition. While Partition gets the best results and gains the most from the samples when there are not many clusters, Exact Dollar Partition gains more with more clusters. This result indicates that if one choose to use Exact Dollar Partition with a large number of clusters, it is even more necessary to increase the number of samples to guarantee better outcome.

\subsection*{\normalsize Varying First Stage Size}
\begin{figure}[ht]
\begin{center}
\includegraphics[width=0.6\linewidth]{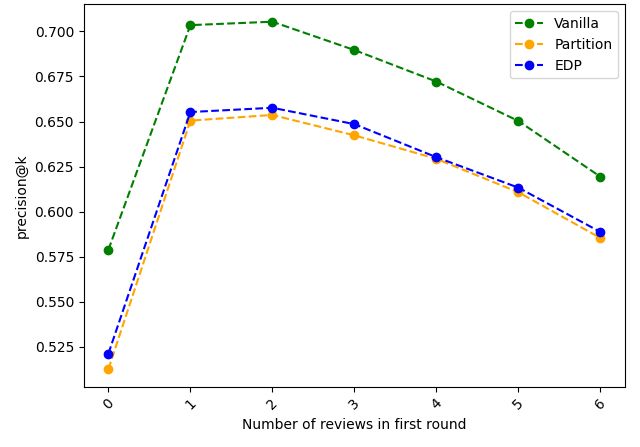}
\caption{Precision@k for 200 items taking top 10 with each agents with 7 overall reviews, and no agents are selected following the first stage and 150 are removed; $\phi=0.95$.
}\label{fig:init_rev}
\Description{Graph presenting precision@k and change in precision@k for for a two stage peer selection instance.}
\end{center}
\vspace{-1.5em}
\end{figure}

The size of the first stage needs to thread a fine line between leaving enough agents to review in the 2nd stage, while having the outcome of the first-stage of high enough quality to allow the removal/acceptance of some papers based on the first stage. As can be seen in Figure~\ref{fig:init_rev}, a small first stage is useful, but the quality decreases as it grows (and, presumably, the second stage cannot make its contribution meaningful). Obviously, the precise location of the optimal number of reviews changes with specific configuration values, but having a first round was beneficial compared to plain single-stage in all cases.


\subsection*{\normalsize How Many Papers to Throw Out?}

\begin{figure}[ht]
\begin{center}
\includegraphics[width=0.7\linewidth]{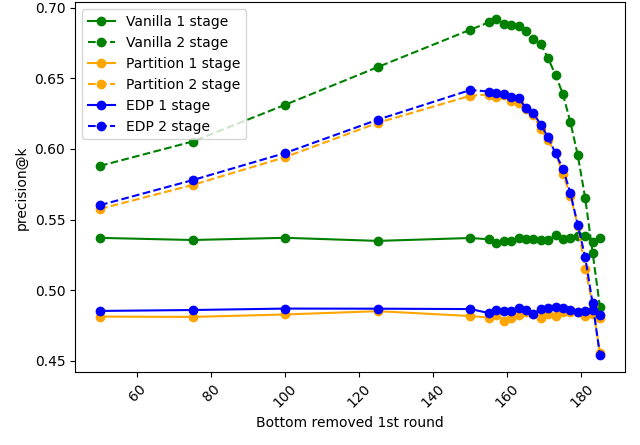}
\caption{Precision@k for 200 items taking top 10 with each agents with 5 overall reviews, of which 1 is in the first stage, and no agents are selected following the first stage; $\phi=0.95$. 
}
\label{fig:chosebottom}
\Description{Graph presenting precision@k and change in precision@k for for a two stage peer selection instance.}
\end{center}
\end{figure}

Intuitively -- and as can be seen from the previous section's results -- removing agents very far from $k$ should have very little influence on the outcome, as they were never really ``in the running'' to be accepted. But as we remove more and more papers in the first stage, we begin to remove agents fairly close to the $k$ cut-off point; these agents are still able to possibly being mistaken as agents that should be chosen. \Cref{fig:chosebottom} shows this sweet-spot, which seems to rely mainly on the information noise and not on the particular algorithm used.

\begin{figure}[ht]
\begin{center}
\includegraphics[width=0.7\linewidth]{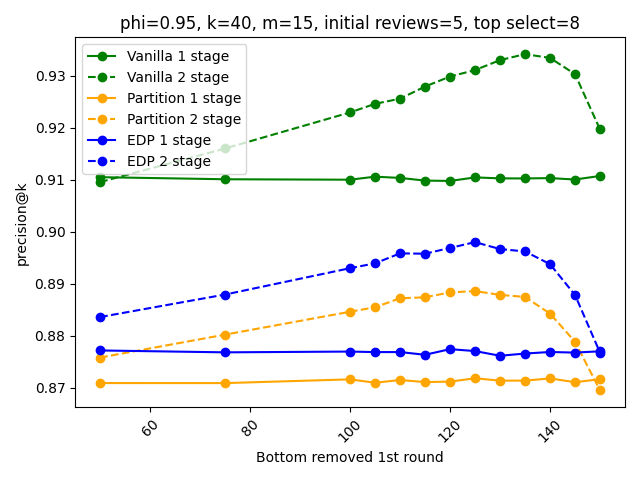}
\caption{Precision@k for 200 items taking top 40 with each agents with 15 overall reviews, of which 8 are in the first stage, and no agents are selected following the first stage; $\phi=0.95$. 
}
\label{fig:chosebottom2}
\Description{Graph presenting precision@k and change in precision@k for for a two stage peer selection instance.}
\end{center}
\end{figure}

Compare this to \Cref{fig:chosebottom2}, displaying the effect of a much richer review environment (each agent reviews much more, allowing for a larger first stage), which one would imagine could allow for a much more exact first stage. However, many more agents are selected ($40$ vs $10$), meaning that the borderline area, where exactness is valued, is a higher value. The value of more reviews in the first stage seems to be more significant, as when $k=10$ (one review in first stage) compares badly to $k=40$ (5 reviews in first stage) -- the precision@$k$ is both much higher, and its optimal value is when more agents are removed following the first stage, i.e., it is more confident in the outcomes of the first stage.

\subsection*{\normalsize How Many Papers to Accept Outright?}

\begin{figure}[ht]
\begin{center}
\includegraphics[width=0.7\linewidth]{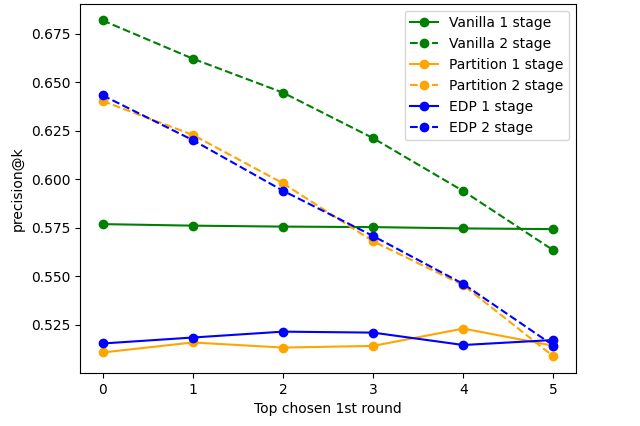}
\caption{Precision@k for 200 items taking top 10 with each agents with 7 overall reviews, of which 3 are reviewed in the first stage, and 100 agents are removed following the first stage; $\phi=0.95$.}\label{fig:chosetop}
\Description{Graph presenting precision@k and change in precision@k for for a two stage peer selection instance.}
\end{center}
\end{figure}

Intuitively, this should be a mirror image of the previous section, as when $N=200$, selecting the top 50 is equivalent to selecting the bottom 150. However, the high noise means that even top agents might not be really the ground-truth top candidate. Thus, rather surprisingly, even adding just one agent to accept after the first stage results in reducing the algorithms' performance. As can be seen in \Cref{fig:chosetop}, while the two-stage mechanism still over-performs the single-stage mechanism, its margins of improving diminish the more agents are selected as top-$k$ following only the first stage.

\section{Discussion}
In this paper, we begin to tackle the issue of two-stage mechanisms in peer evaluation, following it being proposed and implemented since 2020~\cite{LMNZCNR22}. We show that the fundamental assumption of two-stage mechanisms -- that they improve the selection process -- is true, though it can be undermined in the real-world by agents misreporting their preferences. For this reason we also examined two-stage mechanisms in two strategyproof mechanisms, Partition and Exact Dollar Partition, which seem to be more strongly affected by this change. \textbf{We show that in a relatively less noisy environment, the agents benefiting the most from this are the borderline ones, but as the environment gets more noisy, and agents differ more in their views, a two-stage mechanism helps higher ranked agents most.} In every setting we found the two-stage process improves a single-stage one, though the improvement was an order of magnitude more for the strategyproof mechanisms, and in particular Exact Dollar Partition.

Looking into key parameters of the problem, we find that the larger the first round, the more agents can be removed / accepted in the first stage. This is largely unsurprising, as the confidence in the first round's ranking is higher. However, the first stage seems to reach its best contribution when it is significantly smaller than the second one.

%

This exploration of the 2-stage process is only the beginning of the road. Not only because more mechanisms can be examined for their performance, but also because finding techniques to maximize review concentration on the borderline papers is very needed, as the volume of conference submissions grows dramatically, while committees are still relatively small. Are more stages the answer? Perhaps a ``rolling'', continuous mechanism is possible \footnote{Similar to the ARR Rolling Review Process in the NLP community, \url{https://aclrollingreview.org/}}, in which there are no formal stages, but agents review a paper until there are enough reviews for a decision to be reached. Of course, there might be mechanisms for which the two-stage mechanism destroys their strategyproofness or is simply unworkable.

\begin{acks}
Mattei was supported in part by NSF Awards, IIS-RI-2134857, IIS-RI-2339880 and CNS-SCC-2427237 as well as the Harold L. and Heather E. Jurist Center of Excellence for Artificial Intelligence at Tulane University and the Tulane University Center of Excellence for Community-Engaged Artificial Intelligence (CEAI). Portions of this research were conducted with high performance computational resources provided by the Louisiana Optical Network Infrastructure (LONI) (http://www.loni.org).
\end{acks}

%
%




\bibliographystyle{ACM-Reference-Format} 
\bibliography{general}


\end{document}